\documentclass[letterpaper,english,reprint,nofootinbib,aps,superscriptaddress,showpacs,showkeys,prl]{revtex4-1}

\usepackage{babel,calc,amsmath,amsthm,amssymb,graphicx,subfigure,xcolor,comment}
\usepackage{txfonts}
\usepackage{mathdots}
\usepackage{todonotes}
\usepackage[T1]{fontenc}
\usepackage[unicode=true]{hyperref}
\usepackage{booktabs}
\usepackage{threeparttable}
\usepackage{times}
\usepackage{CJK}

\hypersetup{
	colorlinks=true,       		
	linkcolor=blue,          	
	citecolor=blue,            
	urlcolor=black,           	
}

\theoremstyle{definition}

\begin{document}

\begin{CJK*}{UTF8}{gbsn}


\title{Observation of High-Order Quantum Pancharatnam-Berry Phase with Structured Photons}

\author{Shuang-Yin Huang \href{https://orcid.org/0000-0002-5744-5188}}
\email[]{These authors contributed equally to this work.}
\author{He Jiang}
\email[]{These authors contributed equally to this work.}
\author{Zhi-Cheng Ren\href{https://orcid.org/0000-0001-5098-8633}}
\email[]{zcren@nju.edu.cn}
\author{Zi-Mo Cheng}
\author{Wen-Zheng Zhu}
\author{\\Jing Gao}
\affiliation{National Laboratory of Solid State Microstructures and School of Physics, Nanjing University, Nanjing 210093, China}
\affiliation{Collaborative Innovation Center of Advanced Microstructures, Nanjing University, Nanjing 210093, China}

\author{Chang Liu}
\affiliation{Shandong Institute of Quantum Science and Technology Co. Ltd, Jinan, 250101, China}

\author{Xi-Lin Wang\href{https://orcid.org/0000-0002-3990-6454}}
\email[]{xilinwang@nju.edu.cn}
\affiliation{National Laboratory of Solid State Microstructures and School of Physics, Nanjing University, Nanjing 210093, China}
\affiliation{Collaborative Innovation Center of Advanced Microstructures, Nanjing University, Nanjing 210093, China}
\affiliation{Hefei National Laboratory, Hefei 230088, China}
\affiliation{Jiangsu Physical Science Research Center, Nanjing 210093, China}

\author{Hui-Tian Wang\href{https://orcid.org/0000-0002-2070-3446}}
\affiliation{National Laboratory of Solid State Microstructures and School of Physics, Nanjing University, Nanjing 210093, China}
\affiliation{Collaborative Innovation Center of Advanced Microstructures, Nanjing University, Nanjing 210093, China}
\affiliation{Collaborative Innovation Center of Extreme Optics, Shanxi University, Taiyuan 030006, China}

\date{\today}

\begin{abstract}

When a quantum system evolves so that it returns to its initial state, it will acquire a geometric phase acting as a memory of the transformation of a physical system, which has been experimentally measured in a variety of physical systems. In optics, the most prominent example is the Pancharatnam-Berry (PB) phase. Recent technological advances in phase and polarization structure have led to the discovery of high-order PB phases with structured light fields. The study on the high-order PB phase is limited in the context of elementary quantum states of light, especially in the case of photon number states. Here, we experimentally investigate the differences of high-order PB phases between single-photon and N00N states. Our results show that the PB phase, like the dynamic phase, can also be doubled under two-photon states, which can greatly improve the phase sensitivity for greater $N$ in N00N states and high-order structured photons. This may show some implications for quantum precision measurement and quantum state engineering based on geometric phase.
\end{abstract}

\maketitle
\end{CJK*}    

In quantum systems, the phase can be divided into two categories. One is dynamic phase, which can be obtained by solving the time-dependent Schr\"{o}dinger equation or Maxwell equations. The other is geometric phase, first proposed by Berry in 1984 \cite{Berry1984, shapere1989}. The geometric phase acts as a memory of the transformation of a physical system when a quantum system evolves so that it returns to its initial state. It depends on the geometric shape of the quantum parameter space. When the system undergoes simple continuous changes, the phase is related to the curvature of the space \cite{Berry1988,Anandan1992,Cisowski2022,Tiwari2007}. Since the introduction of geometric phase, it has been experimentally measured in a variety of physical systems including optical physics \cite{Chyba1988,Tomita1986,Tiwari2007,Jisha2021,Karniel2019}, microwave \cite{Leek2007,Ralston1998}, ultra-cold atoms \cite{Bharath2019}, sound waves \cite{Xiao2015},  trapped ions \cite{Leibfried2003}, superconducting nanocircuits \cite{Falci2000}, chemical reactions \cite{Clary2005}, condensed matter \cite{Dutreix2019}, etc. Due to the geometric phase depends on the geometric shape of the parameter space and is significantly different from dynamic phase, it is easier to modulate and control. Therefore, it has extensive attention and research in fields such as quantum computing \cite{duan2001,Nigg2014}, quantum state engineering \cite{Sjoqvist2015,Jha2008}, quantum precision measurement \cite{Cho2019}, frequency shift \cite{Simon1988,Cheng2023} and artificial material design \cite{Litchinitser2016,Devlin2017,Slussarenko2016}. 

In optical systems, the geometric phase that arises from evolution in the polarization state (that is associated with spin angular momentum, SAM) was first discovered by Pancharatnam in 1956 \cite{Pancharatnam1956}. Berry's result was the generalization of an earlier result of Pancharatnam. Therefore, such a geometric phase is also known as the Pancharatnam-Berry (PB) phase in optics. The PB phase is acquired when a polarization state undergoes a closed trajectory on the Poincar\'{e} sphere (PS), which differs markedly from a dynamic phase being exceedingly more robust \cite{Berry1987}. In recent years, PS has been extended to various types by introducing new degrees of freedom beyond polarization to represent more general quantum states, which has greatly promoted their applications in quantum information and the control of light field \cite{Wang2015,Kong2019,Ni2021}. For example, Padgett \textit{et al.} replaced polarization with orbital angular momentum (OAM) and created a mode sphere of OAM to explain the evolution of transverse modes \cite{Padgett1999}. The experimental observation of the PB phase in mode transformations on the OAM mode sphere be also reported and the PB phase is proportional to the OAM \cite{Galvez2003}. Further, Milione \textit{et al}. develop a high-order PS to describe the vector vortex light fields and the PB phase on the high-order PS has been demonstrated to be proportional to the total angular momentum (TAM) \cite{Milione2011,Milione2012}. Based on the mode transformations involving spatial mode and polarization with SU(2) symmetry, Shen \textit{et al.} proposed a generalized SU(2) PS for general high-dimensional structured light field \cite{Shen2020,Cisowski2022}.

Although the geometric phase is a common physical phenomenon in optical systems, the study is limited to the classical field and single-photon level \cite{Kwiat1991}. The PB phase with transformation of the polarization has be mentioned in some two-photon interference experiments \cite{Brendel1995,Kobayashi2011,Qi2020}. While, the study on the high-order PB phase is limited in the context of elementary quantum states of light, especially in the case of photon number states. In general, any phase acquired by a mode of a quantum system leads to a photon-number dependent phase for the quantum state. When $N$ photons occupy the same mode, the quantum state obtains $N$ times the same phase \cite{Hiekkamaki2021,Hiekkamaki2022}. Is it equally effective for the geometric phase? We all know that the geometric phase is different from dynamic phase and depends only on the geometric shape of the parameter space. 

In this work, we theoretically and experimentally describe the high-order quantum PB phase based on single-photon and two-photon states. We find that the PB phase is the same as the dynamic phase and will be doubled under the N00N state with $N = 2$. The high-order quantum PB phase under the measuring of the N00N state can greatly improve the phase sensitivity with high-order structured photons. This may show some implications for quantum precision measurement and quantum state engineering based on the geometric phase.

\begin{figure}[bhtp]
	\centering
	\includegraphics[width=0.95\linewidth]{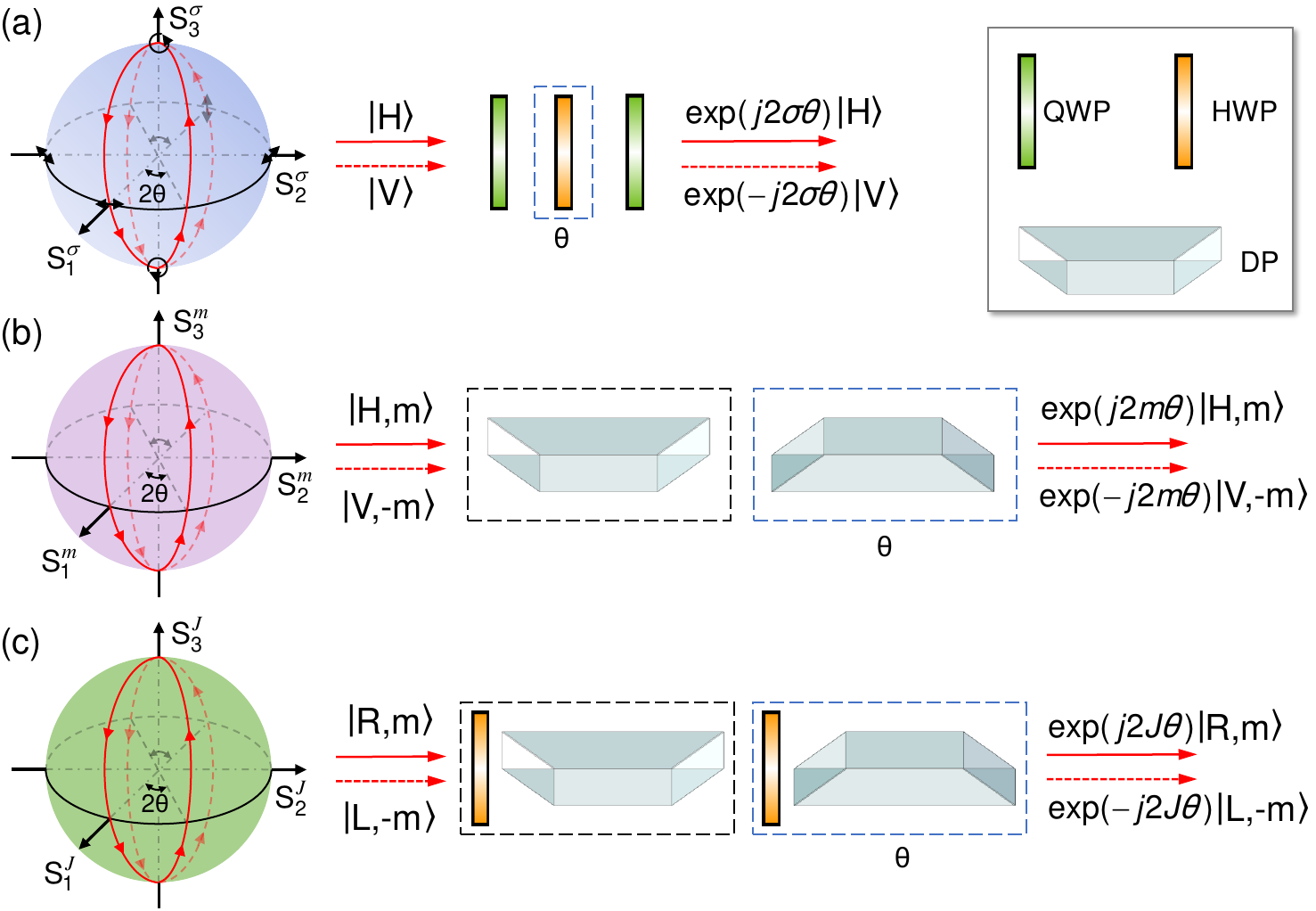}
	\caption{\label{fig:1}The geometric phase acquired with the transformations of polarization, transverse mode, and transverse vector mode can be described by a generalized SU(2) PS.
		(a) Closed trajectory of the polarization state evolution on the PS and control unit of the PB phase.
		(b) Closed trajectory of the OAM mode evolution on the OAM mode sphere and control unit of the geometric phase.
		(c) Closed trajectory of the vector mode evolution on the high-order PS and control unit of the high-order PB phase.
		$|H\rangle$ and $|V\rangle$ represent horizontally and vertically polarized states, respectively. $|R\rangle$ and $|L\rangle$ stand for right and left circularly polarized states, respectively. $\sigma=\pm{1}$ indicate the spin angular momentum (SAM) of the $|R\rangle$ and $|L\rangle$ photons, respectively.
		$\left| \pm{m} \rangle \right.$ are the OAM eigenstates with the vortex phase $\exp{(\pm{jm\phi})}$. $|P,m\rangle$ stands for the $|P\rangle$ polarized OAM eigenstate with the vortex phase $\exp{(jm\phi)}$, where $|P\rangle$ can be $|H\rangle$, $|V\rangle$, $|R\rangle$ or $|L\rangle$, especially $|P,0\rangle$ is the $|P\rangle$ polarized fundamental Gaussian mode with $m = 0$.
		HWP, half-wave plate; QWP, quarter-wave plate; DP, Dove prism.
	}  
\end{figure}

The transformations based on polarization, transverse mode, and transverse vector mode can all be described by a generalized SU(2) PS \cite{Shen2020}. In the parameter space of the generalized SU(2) PS, the north and south poles are always chosen to be two orthogonal bases. For example, as shown in Fig.~\ref{fig:1}(a), the initial $|H\rangle$ ($|V\rangle$) state on the equator of the standard PS experiences a cyclic evolution along the anticlockwise red-solid-line (clockwise red-dashed-line) trajectory surrounding a solid angle $\Omega = 4\theta$ on the PS, which can be achieved by a control unit composed of a HWP (at a rotation angle of $\theta$) sandwiched between two QWPs, thus a PB phase of $\Phi^{SAM} = 2\sigma\theta$ ($-2\sigma\theta$) will be acquired, where $\sigma=\pm{1}$ and it indicates the spin angular momentum (SAM) of the $|R\rangle$ and $|L\rangle$ photons, respectively. The PB phase is a half of $\Omega$. As shown in Fig.~\ref{fig:1}(b), after passing through two DPs (with a relative orientation angle of $\theta$), the initial OAM state $|H,m\rangle$ ($|V,-m\rangle$)  experiences a cyclic evolution along the anticlockwise red-solid-line (clockwise red-dashed-line) trajectory, which surrounds a solid angle $\Omega = 4\theta$ on the mode sphere, a geometric phase of $\Phi^{OAM} = 2m\theta$ $(-2m\theta)$ can be acquired. As shown in Fig.~\ref{fig:1}(c), the state $|R,m\rangle$ $(|L,-m\rangle)$ experiences a cyclic evolution along the anticlockwise red-solid-line (clockwise red-dashed-line) trajectory on the high-order PS to return to its initial state, which can be achieved through two sets of HWP+DP, in which the second set of HWP+DP has a rotation angle of $\theta$ with respect to the first one. The cyclic trajectory on the high-order PS surrounds a solid angle $\Omega = 4\theta$, thus the state $|R,m\rangle$ ($|L,-m\rangle$) undergoing a cyclic evolution will introduce a high-order PB phase of $\Phi^{TAM} = 2 J \theta$ ($- 2 J \theta$), where $J = m + \sigma$ is the TAM. 

When a single photon at the initial state $|\psi_1\rangle$ undergoes a cyclic transformation on the high-order PS to be converted into a series of new states and finally returns to its initial state $|\psi_1 '\rangle$, it will acquire a high-order PB phase given by
\begin{equation}
|\psi_1'\rangle = e^{j\rm{arg}\langle\psi_1'|\psi_1\rangle}|\psi_1\rangle = e^{-j J \Omega/ 2}a^{\dagger}|0\rangle = e^{-j(m+\sigma)\Omega / 2}a^{\dagger}|0\rangle.
\end{equation}
Here, $|0\rangle$ indicates vacuum state and $a^{\dagger}$ is the creation operator of photon. While, for the $N$ identical photons in the initial state $|\Psi_1\rangle$, it also undergoes a cyclic transformation and also returns to its initial state $|\Psi_1 '\rangle$. The $N$ photons can be expressed as the $N$th tensor product and the high-order PB phase for the evolution of an $N$-photon Fock state can be written as
\begin{equation}
|\Psi_1'\rangle^{\otimes N}=e^{jN\rm{arg}\langle\Psi_1'|\Psi_1\rangle}|\Psi_1\rangle^{\otimes N}= \frac{e^{-jN(m+\sigma)\Omega / 2}}{\sqrt{N!}}(a^{\dagger})^N|0\rangle,
\end{equation}
where $\otimes$ represents the tensor product. So, $N$-photon Fock state is expected to acquire $N$ times the geometric phase for one photon, showing the same result as the dynamic phase.

\begin{figure*}
    \centering \includegraphics[width=0.95\linewidth]{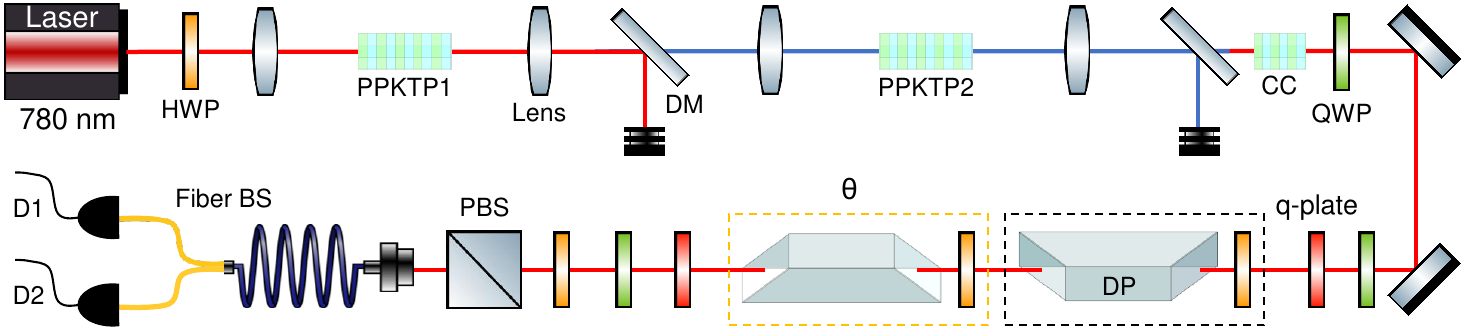}  
      \caption{\label{fig:2}Experimental setup for measuring the high-order PB phase with the N00N state. A Gaussian beam at $\lambda = 780$ nm is doubled to 390 nm by using periodically poled potassium titanyl phosphate crystal (PPKTP1) crystal. A linearly polarized 390 nm beam is incident into type-II PPKTP2 to create a pair of collinear photons via the SPDC process. DM is the dichroic mirror to isolate the pump beam. The extra birefringent crystals CC is used to compensate for the birefringent walk-off effects from the PPKTP2. Both PPKTP crystals have a period of 7.825 $ \mu $m and a length of 10 mm. The compensating crystal is a KTP crystal with a length of 5 mm. Passing through a QWP at 45$^{\circ }$, the two photons will be transformed into the N00N state in the basis of $|H\rangle$ and $|V\rangle$. The QWP and q-plate create high-order structured photons. The control unit of high-order PB phase is composed of two HWPs and two Dove prisms (HWP+DPs). The second q-plate converts structured photons into fundamental mode. Photons are projected into the basis of $|D\rangle$ after passing through an HWP at 22.5$^{\circ }$ and a polarization beam splitter (PBS). Then, photons are collected by a $1 \times 2$ single-mode optical fiber beam splitter (BS). The two photons are detected by two single-photon avalanche photodiodes (D1 and D2) without cooling and with the dark count rate of < 200 cps.}
\end{figure*}

In our experimental setup shown in Fig.~\ref{fig:2}, a continuous-wave laser with a power of 2 W, a wavelength of $\lambda = 780$ nm, and a narrow linewidth of 50 kHz is converted to its second harmonic at 390 nm by frequency doubling in a type-II PPKTP1 crystal at a phase-matching temperature of 26.2$^\circ \rm{C}$. The generated linearly polarized second harmonic at 390 nm pumps PPKTP2 to create a pair of collinear photons via the spontaneous parametric down-conversion (SPDC) process. The PPKTP2 is designed to be identical to the PPKTP1 with a length of 10 mm kept at 25.6$^\circ \rm{C}$ for the degenerate wavelength of 780 nm. After compensating the birefringent walk-off effect and passing through a QWP at 45$^{\circ }$, the two photons in $|H\rangle$ and $|V\rangle$ will be transformed into the N00N state in polarization \cite{SuppMat}
\begin{equation}
|\Psi\rangle=\frac{1}{\sqrt{2}}\left(|2\rangle_H|0\rangle_V+|0\rangle_H|2\rangle_V\right) . 
\end{equation}
For the N00N state passing through the QWP and q-plate, the two-photon in $|H\rangle$ and $|V\rangle$ will be prepared into the states $\vert R,m\rangle$ and $\vert L,-m\rangle$, represented by the north and south poles of high-order PS, respectively. To experimentally introduce the high-order PB phase for $\vert R,m\rangle$ and $\vert L,-m\rangle$, a cyclic transformation of state on the high-order PS is physically carried out, independently. The paths considered are the geodesic loops (red solid lines and red dashed lines between two poles) shown in Fig.~\ref{fig:1}(c). The control unit containing two sets of HWP+DP that transform both the SAM and OAM are employed, as shown by the two dashed boxes in Figs.~\ref{fig:1}(c) and~\ref{fig:2}. The two sets of HWP+DP will transform into the initial state completing the path forming a closed loop on the high-order PS. It will acquire an additional high-order PB phase depending on the relative rotation angle $\theta$ between the two sets of HWP+DP. A high-order PB phase of $\pm2N(m+\sigma)\theta$ will be acquired for the incident $N$ photons in $\vert R,m\rangle$ and $\vert L,-m\rangle$. To measure the high-order PB phase, we use a second q-plate to do the reverse transformation induced by the first q-plate. In this way, the structured
OAM photons are transformed into the fundamental Gaussian mode. Passing a QWP, the high-order PB phase will be kept in the relative phase between the $|H,0\rangle$ and $|V,0\rangle$ states. Then, a HWP at 22.5$^{\circ }$ and a PBS will help us to extract the relative phase \cite{SuppMat}. For the N00N state with $N = 2$, we use a $1 \times 2$ fiber BS (50:50) to detect the coincidence count of D1 and D2. In experiment, an additional QWP is added on the DP to compensate for the change of polarization. Due to the beam deviation caused by the rotation of the DP, we need to realign the center of the q-plate and adjust the coupling of the fiber BS to obtain the same maximum coincidence count for each measurement.

\begin{figure*}
    \centering  \includegraphics[width=0.93\linewidth]{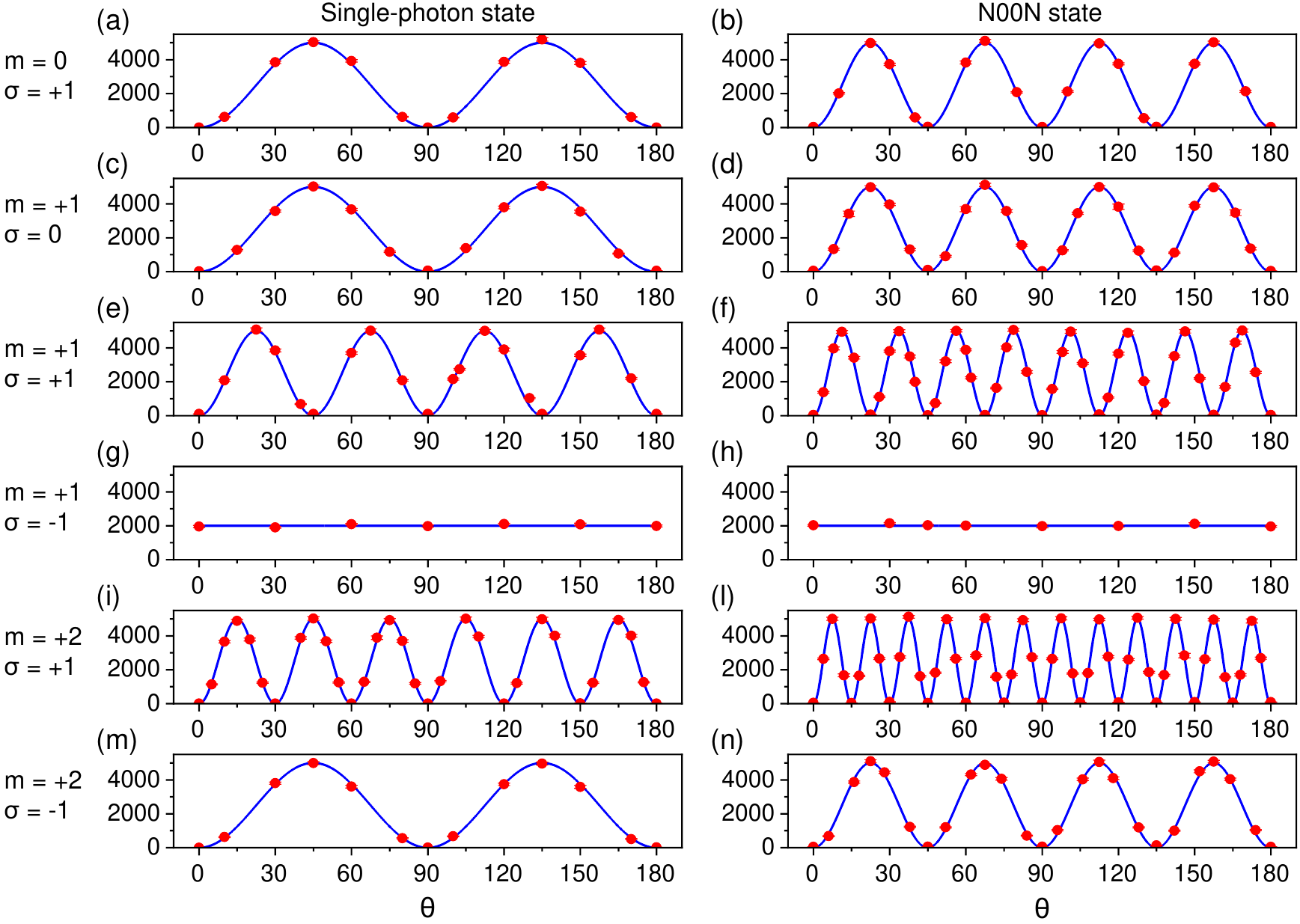}   
    \caption{\label{fig:3}  Contrasting the single photon state (left column) with the N00N state (right column) for the geometric phase with different order. The theoretical (blue solid lines) and experimental (red dots) coincidence count as a function of relative rotation angle $\theta$.  (a) and (b) show the PB phase of polarization evolution and the coincidence count is a cosine function of rotation angle $\theta$ of HWP. (c) and (d) show the geometric phase based on OAM mode with $\sigma = 0$ and the coincidence count is a cosine function of rotation angle $\theta$ of Dove prism. (e-h) show the high-order PB phase based on vector modes with $m =+1, +2$ and $\sigma = \pm{1}$. The coincidence count is a cosine function of rotation angle $\theta$ of DP+HWP. }
\end{figure*}

The PB phase related to polarization is measured as shown in Fig.~\ref{fig:3}(a, b). The PB phase can be acquired by a control unit composed of a HWP sandwiched between two QWPs, we set two QWPs at 45$^{\circ }$ and HWP at angle $\theta$ to acquire the controllable PB phase of $2\sigma\theta$ ($-2\sigma\theta$) for incident photon in $|H\rangle$ ($|V\rangle$). For a single photon in $|D\rangle = \frac{1}{\sqrt{2}} (|H\rangle + |V\rangle)$, the different geometric phase will be acquired for $|H\rangle$ and $|V\rangle$ and it ultimately becomes $\exp{(j2\sigma\theta)} |H\rangle+\exp{(-j2\sigma\theta)}|V\rangle$ \cite{SuppMat}. After projection to the basis of $|A\rangle = \frac{1}{\sqrt{2}} (|H\rangle - |V\rangle)$ to get the minimum coincidence count at the beginning of the measurement, the coincidence count for the
single-photon state is proportional to $1 - \cos(4\sigma\theta)$, as shown by the blue curve in Fig.~\ref{fig:3}(a), and the experimentally measured coincidence counts (red dots) varies with the rotation angle $\theta$ in a cosine curve with a period of 90$^{\circ}$. While, for the N00N state with $N = 2$, the curve still keeps a cosine profile but the period reduces to the half being 45$^{\circ }$ shown in Fig.~\ref{fig:3}(b).

To explore the geometric phase acquired by spatial OAM mode, we use a q-plate sandwiched between two QWPs to prepare the states $|H,m\rangle$ and $|V,-m\rangle$ from the $|H,0\rangle$ and $|V,0\rangle$ modes, respectively. After passing through two DPs, the OAM modes return to their respective initial states $|H,m\rangle$ and $|V,-m\rangle$. The geometric phase generated by the evolution of OAM modes carried by orthogonal polarization states is proportional to the relative rotation angle $\theta $ of two DPs as $\exp{(j2m\theta)} |H,m\rangle + \exp{(-j2m\theta)}|V,-m\rangle$ \cite{SuppMat}. After passing through the second q-plate sandwiched between two QWPs, the OAM modes will be transformed into fundamental Gaussian mode and the geometric phase will be kept in the relative phase of $|H,0\rangle$ and $|V,0\rangle$. Converting into the basis of $|D,0\rangle$ and setting the initial angle of the first DP at $\theta_0 = \pi/4m$, we can obtain that the coincidence count is proportional to $1-\cos(4m\theta)$ for the single photon state. For $m=1$, the theoretical coincidence count varies with $\theta$ in a cosine curve is shown in the blue curve of Fig.~\ref{fig:3}(c). In experiment, the coincidence count (red dots) varies with $\theta$ with a period of 90$^{\circ}$. While, for the N00N state with $N = 2$, the period decreases to the half (i.e. 45$^{\circ }$) shown in Fig.~\ref{fig:3}(d). For $m=2$ shown in Fig. S2 in supplementary materials, the coincidence count of single photon varies with a period of 45$^{\circ}$, but for the N00N state with $N=2$, the period becomes 22.5$^{\circ}$. We find that both $N$ and $m$ can affect the geometric phase and the larger $m$ and $N$, the more sensitive the phase change is.
  
The high-order PB phase generated by the vector mode can be described by the high-order PS. We use a QWP and a q-plate to prepare the initial vector modes $|R,m\rangle$ and $|L,-m\rangle$ from $|H,0\rangle$ and $|V,0\rangle$, respectively. After passing through the control unit composed of two sets of HWP+DP, the vector modes $|R,m\rangle$ and $|L,-m\rangle$ experience a cyclic evolution and then return into the initial states, correspondingly, the high-order PB phases of $2(m+\sigma)\theta$ and $-2(m+\sigma)\theta$ are acquired, depending on the relative rotation angle $\theta$ between two sets of HWP+DP \cite{SuppMat}. To easily measure the high-order PB phases, we need to convert the vector modes into $|H,0\rangle$ and $|V,0\rangle$ by a q-plate and a QWP, because the high-order PB phases are kept in the relative phase of $|H,0\rangle$ and $|V,0\rangle$. Converting into the basis of $|D,0\rangle$ and setting the initial angle of the first set of HWP+DP at $\theta_0 = \pi / 4(m + \sigma)$, we can obtain the coincidence count to be $1 - \cos [4 (m + \sigma) \theta]$ for the single-photon state. For the case of $(m,\sigma) = (+1,+1)$ shown in Fig.~\ref{fig:3}(e, f), the coincidence count of the single photon state varies with a period of 45$^{\circ}$, but for the N00N state, the period becomes 22.5$^{\circ }$. For the case of $(m,\sigma) = (+1,-1)$ shown in Fig.~\ref{fig:3}(g, h), the coincidence counts of the single photon and N00N state remain unchanged with $\theta$ and the high-order PB phase vanishes. For the case of $(m,\sigma) = (+2,+1)$ shown in Fig.~\ref{fig:3}(i, l), the coincidence count for the single-photon state changes with $\theta$ in a period of 30$^{\circ }$ and the period becomes 15$^{\circ }$ for the N00N state. For the case of $(m,\sigma) = (+2,-1)$ shown in Fig.~\ref{fig:3}(m, n), the coincidence count for the single-photon state changes with $\theta$ in a period of 90$^{\circ}$ and the period becomes 45$^{\circ }$ for the N00N state. These are in good agreement with the high-order PB phase in theory, where the obtained geometric phase is proportional to the TAM of $J = m + \sigma$.
\begin{figure}
    \centering \includegraphics[width=\linewidth]{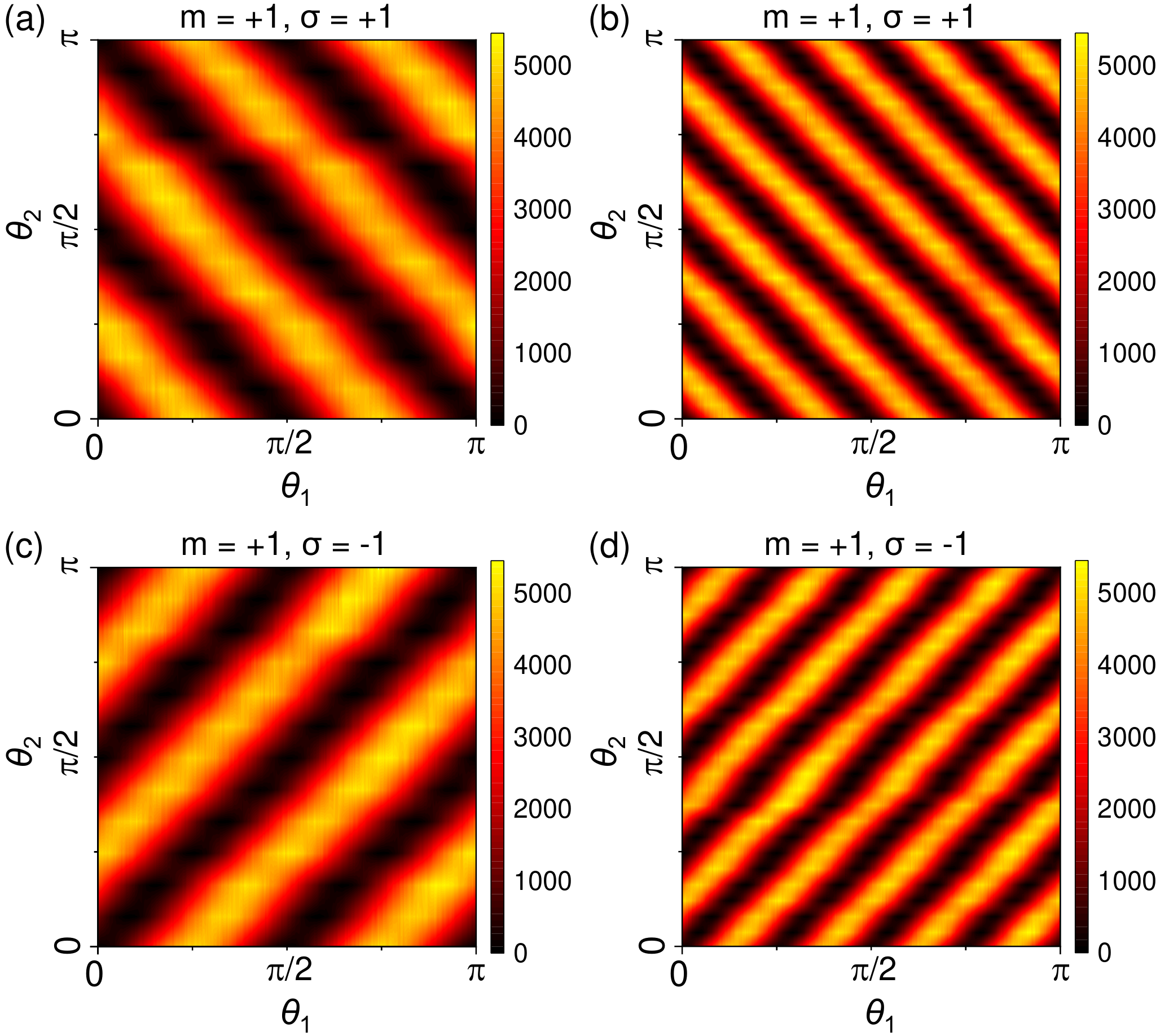}   
    \caption{\label{fig:4} Contrasting the single-photon state (left column) with the N00N state (right column) for high-order PB phase in SAM and OAM. The coincidence counts as a function of $\theta_1$ and $\theta_2$, where $\theta_1$ is the relative rotation angle of two HWPs and $\theta_2$ is the relative rotation angle of two DPs. (a) and (b) show the mode in the case of $(m,\sigma) = (+1,+1)$. (c) and (d) show the mode in the case of $(m,\sigma) = (+1,-1)$. The average visibility in (a)-(d) is 0.99, 0.98, 0.99, and 0.98, respectively.}
\end{figure}

The high-order PS can be seen as a combination of the standard PS and the OAM mode sphere. Due to the independence of SAM and OAM, the high-order PB phase can be derived from the geometric phases acquired by the separate evolution of SAM and OAM modes. For the single photon in the mode of $(m,\sigma) = (+1,+1)$ and setting the initial angle of the first set of HWP+DP at $\theta_0 = \pi / 8$, the coincidence count is given by $1 - \cos (4\theta_1+4\theta_2)$, where $\theta_1$ ($\theta_2$) is the relative rotation angle of two HWPs (two DPs). The function $1 - \cos (4\theta_1+4\theta_2)$ can form a two-dimensional graph as shown in Fig.~\ref{fig:4}(a). The parallel stripes are oriented towards $-45^{\circ}$, with a period of 90$^{\circ}$ in both $\theta_1$ and $\theta_2$ dimensions. For the N00N state with $N=2$, the parallel stripes are also oriented towards $-45^{\circ}$, but the periods become 45$^{\circ}$ in both $\theta_1$ and $\theta_2$ dimensions and are half of the single-photon state, as shown in Fig.~\ref{fig:4}(b). For the single photon in the mode of $(m,\sigma) = (+1,-1)$ shown in Fig.~\ref{fig:4}(c), the orientation of the parallel stripes changes to $+45^{\circ}$, but the periods remain 90$^{\circ }$ in both $\theta_1$ and $\theta_2$ dimensions. As shown in Fig.~\ref{fig:4}(d), the N00N state is similar to the single-photon state in Fig.~\ref{fig:4}(c) except that the period of stripes becomes half. Since the high-order PB phase is a linear superposition of the geometric phases from SAM and OAM, these two parts are independent of each other. For the N00N state, the geometric phase is doubled in both polarization and spatial degrees of freedom.

Generally speaking, the geometric phase is different from the dynamic phase as it solely depends on the geometric shape of the parameter space and the closed geodesic surface formed during the state's evolution. It is independent of the number of photons in the evolution process. This gives us a misunderstanding that the geometric phase does not change under the measurement of $N$-photon Fock state. Through our research on the N00N states, we find that $N$ times the geometric phase can be acquired for $N$ identical photons. This also means that the N00N states cannot be simply described as a phenomenon that the wavelength is $\lambda/N$. Besides, for the N00N state, the estimated quantum Fisher information $F_Q = N^2(m+\sigma)^2$ is consistent with the well-known Heisenberg limit that the lower bound of the standard deviation for measuring the geometric phase is proportional to $1 / [N(m+\sigma)]$ \cite{SuppMat}. Hence, using the N00N state with larger $N$ should be able to enhance the sensitivity of quantum PB phase. We also know that the so-called high-order PS is simply a combination of the standard PS and the mode sphere. The high-order PB phase is actually a linear superposition of the geometric phase in SAM and OAM, which can be independently controlled. In particular, the OAM is easily controlled, so increasing the topological charge $m$ can also effectively enhance the sensitivity. 

In summary, we have confirmed both theoretically and experimentally the geometric phase based on polarization and spatial mode. The high-order PB phase is actually a linear superposition of geometric phase in SAM and OAM. The N00N states can acquire $N$ times the geometric phase of the single photon state. Increasing the number of photons $N$ and topological charge $m$ can improve the resolution of measuring the phase~\cite{Hiekkamaki2021, Hong2023}. This has an important inspiration and assistance for quantum precise measurement and quantum state engineering based on geometric phase.

~\
\begin{acknowledgments}
This work was supported financially by the National Natural Science Foundation of China (12304359, 12404382, 12234009, and 12274215); the National Key R\&D Program of China (2020YFA0309500); the Innovation Program for Quantum Science and Technology (2021ZD0301400); the Program for Innovative Talents and Entrepreneurs in Jiangsu; the Natural Science Foundation of Jiangsu Province (BK20220759); the Shandong Provincial Natural Science Foundation (ZR2021LLZ010); the Key R\&D Program of Guangdong Province (2020B0303010001); the China Postdoctoral Science Foundation (2023M731611); the Jiangsu Funding Program for Excellent Postdoctoral Talent (2023ZB717); the Key R\&D Program of Jiangsu Province (BE2023002); the Natural Science Foundation of Jiangsu Province (BK20233001).
\end{acknowledgments}


\end{document}